# Femtosecond electrons probing currents and atomic structure in nanomaterials

Melanie Müller, Alexander Paarmann, and Ralph Ernstorfer

*Fritz-Haber-Institut der Max-Planck-Gesellschaft, Faradayweg 4-6, D-14195 Berlin, Germany*

## Abstract

The investigation of ultrafast electronic and structural dynamics in low-dimensional systems like nanowires and two-dimensional materials requires femtosecond probes providing high spatial resolution and strong interaction with small volume samples. Low-energy electrons exhibit large scattering cross sections and high sensitivity to electric fields, but their pronounced dispersion during propagation in vacuum so far prevented their use as femtosecond probe pulses in time-resolved experiments. Employing a laser-triggered point-like source of either divergent or collimated electron wave packets, we developed a hybrid approach for femtosecond point projection microscopy and femtosecond low-energy electron diffraction. We investigate ultrafast electric currents in nanowires with sub-100 femtosecond temporal and few 10 nm spatial resolutions and demonstrate the potential of our approach for studying structural dynamics in crystalline single-layer materials.

## Introduction

One- and two-dimensional crystalline materials have emerged as fundamental building blocks for nanoscale devices[1–3]. Compared to the respective bulk materials, the reduced dimensionality of the translational symmetry has profound effects on the ground state properties of nanomaterials as well as on the coupling between electronic, nuclear and spin degrees of freedom, dictating the dynamical behavior. As all devices operate in states out of equilibrium, and as the dwell time of excited electrons in nanostructures is comparable to the time scale of typical relaxation processes, electron-lattice-spin interactions crucially determine the functionality of future nanodevices. A range of ultrafast laser-based techniques is nowadays available for probing the evolution of electronic, optical, structural and magnetic properties of solids after a sudden perturbation like optical excitation, providing invaluable information on the mutual coupling of electronic, nuclear and spin degrees of freedom as well as of transport properties. Despite femtosecond temporal resolution, the investigation of ultrafast processes in nanoscaled, low-dimensional systems additionally requires high spatial resolution[4–6] as well as high sensitivity sufficient for investigating small sample volumes, i.e., femtosecond probe pulses strongly interacting with the sample. Electrons with sub-keV kinetic energies, here referred to as low-energy electrons, exhibit exceptionally high scattering cross section and a de Broglie wavelength on the order of 1 Å, which, in principle, allows for achieving atomic resolution both in imaging as well as diffraction approaches. Whereas the spatial resolution of current techniques for time-resolved nanoscale imaging of electric fields relies on the near field enhancement at nanostructures[5,6], the high sensitivity of low-energy electrons to electric fields further permits the investigation of weak field distributions in the vicinity of nanoobjects[7]. While the generation of few-femtosecond electron pulses is readily achieved by photoemission[8–11], the biggest challenge in using low-energy electrons as ultrafast probe is to maintain femtosecond duration of the electron pulses during delivery to the sample.

Unlike optical laser pulses, femtosecond electron pulses suffer from temporal broadening in vacuum during propagation to the sample, especially at low energies[12]. Many-electron pulses can be strongly affected by space charge broadening due to Coulomb repulsion[13]. Furthermore, even single electron wave packets experience significant dispersive broadening depending on their initial energy distribution[14]. Temporal compression techniques can be used to obtain femtosecond many-electron pulses at a distant sample[15], but have yet to be demonstrated for low electron energies. Alternatively, space charge broadening can be eliminated by using single electron pulses at high repetition rates[16,17]. Still, achieving femtosecond time resolution with dispersing sub-keV single electron pulses further requires considerable reduction of the propagation distances[18,19]. In our approach, we accomplish femtosecond time resolution by minimizing the electron propagation length down to the μm-range in combination with using single electron pulses. We developed a compact hybrid approach for femtosecond low-energy electron diffraction (fsLEED) and femtosecond point projection microscopy (fsPPM) with electron energies in the range 20 to 1000 eV. A laser-triggered metal nanotip provides a compact point-like source of coherent femtosecond electron wave packets[8–11], optionally collimated for diffraction or spatially diverging for microscopy[7,19,20]. Employing the microscopy mode of operation, we investigate ultrafast currents in axially doped InP nanowires (NWs) with femtosecond temporal and nm spatial resolution. The potential of the diffraction mode to study ultrafast structural dynamics in two-dimensional materials is demonstrated by recording high-quality diffraction images of single-layer graphene with femtosecond electron pulses.

## Results

Figures 1a) and 1b) show the two operation modes for fsPPM and fsLEED, respectively. A tungsten nanotip is positioned at sub-mm distances in front of the sample. Photoelectrons are generated by focusing an ultrashort laser pulse on the negatively biased tip and are accelerated towards the grounded sample. For time-resolved pump-probe experiments, a second laser pulse is focused on the sample under 45° and the arrival time between the two pulses can be varied with an optical delay stage. Projection images and diffraction patterns are recorded with a microchannel plate (MCP) as electron detector positioned 10 cm behind the sample (more details on the setup are described in the Methods and in the Supplementary Section I).

For collimation and energy tuning, we place the tip inside an electrostatic microlens, being either directly coated onto the shaft of the tip[21] or using a metal-coated ceramic microtube. Examples of the potential and electric field $E_z$ in the vicinity of the tip's apex for the imaging and diffraction mode are plotted in Figures 1c) and 1d), respectively. The electric field strength at the apex can be adjusted via the lens voltage independent of the tip voltage, enabling energy tuning at a constant emission current[21]. For diffraction, the electron beam is collimated by flattening the potential field lines around the apex. This is accompanied by a reduction of DC field enhancement, and no field emission is possible in the diffraction mode. However, the nanotip still enhances the optical laser field, leading to localized photoemission from the apex[22]. The photoemission process at the tip is characterized by measuring an interferometric autocorrelation of the photocurrent with the tip as nonlinear medium[23], as plotted in Figure 1e). The peak/baseline ratio of 27:1 reveals a 3$^{rd}$ order emission process, implying that the electron emission is temporally confined to ~3 fs in case the tip is illuminated with 5 fs-laser pulses (laser system described in the methods summary).

**Femtosecond point projection microscopy**

We performed fsPPM measurements on axially doped p-i-n InP nanowires[24] with a 60 nm long i-segment in the center, spanning across 2 μm holes in a gold substrate, see Figure 2a). A projection image of a single NW recorded in field emission mode at a distance $d = 20$ μm and at 90 eV electron energy is shown in Figure 2b). Noticeably, the wire diameter appears bright and much larger than its projected real space diameter. Due to the low electron energies, the projection image is in fact not a shadow image of the spatial shape of the nanoobject, but is rather revealing the local electrostatic field in the objects near-surface region deflecting the electron trajectories[7,25]. These static lensing effects critically depend on extrinsic parameters such as the tip field[7,26], and intrinsic parameters like work function variations, e.g. between the NW and the substrate.

Furthermore, we observe a step of the projected NW diameter $d_{NW}$ close to the NW center (the detailed analysis can be found in the Supplementary Section II.a). Figure 2c) shows line profiles through the NW at two different positions along the wire, revealing a difference of $d_{NW,1} - d_{NW,2} \approx 60$ nm in the projected sample plane. This contrast can be explained by different electric fields surrounding the NW induced by spatial variations of the work function. Numerical simulations show that the observed step corresponds to a difference of the local potential in the 100 meV range, and a difference in the radial electric field around the NW on the order of a few MV m$^{-1}$ (more details on the simulations can be found in the Supplementary Section III.a). In general, the homogeneity of the projected width of a NW with constant radius depends on its specific surface condition, i.e. its doping level, crystal structure and chemical composition[27–29].

The transient change of the NW diameter $\Delta d_{NW}$ after fs laser excitation is plotted in figure 3c) for both segments along the NW, indicated by the two lines in figure 3a). At temporal overlap we observe a clear pump-induced, spatially inhomogeneous change of $d_{NW}$, which axially varies along the NW, as apparent in the difference image taken at 150 fs in Figure 3b). We observe a difference in the maximum amplitudes of the transient signal of $\Delta d_{NW,1}^{max} \approx 5 \cdot \Delta d_{NW,2}^{max}$ for the two segments. Both transients have a fast initial rise, followed by a multi-exponential decay on the femtosecond to picosecond time scale.

In addition to the intentional axial doping, we expect the NWs to exhibit an effective radial doping induced by surface states, pinning the Fermi level and leading to band bending far into the NW[28], as sketched in Figure 3d). The associated surface-space-charge field strongly differs for the different doping types, being larger for the p- than for the n-doped segment[28]. In particular, the effective radial doping profile of the p-segment changes from p-doping in the NW bulk to n-doping at the NW surface, whereas the n-segment exhibits a radial n-n$^+$-profile, leading to an effective n-n+ doping axially along the NW surface, which reduces the axial doping contrast without photoexcitation. After above-bandgap photoexcitation, electrons and holes homogeneously generated in the NW bulk are radially separated by the surface field, leading to radial photocurrents $j_e$ and $j_h$, respectively. This carrier separation, however, transiently reduces the surface band bending due to screening of the space-charge fields[30], leading to a transient shift of the vacuum level, indicated by the red shaded area in Figure 3d). As this is accompanied by a change $\Delta E_{NW}$ of the local electric field at the NW surface, we can monitor these shifts by a transient change of the projected NW diameter being directly proportional to $\Delta E_{NW}$ (see Supplementary Fig. S5). Consequently, the derivative $\Delta d_{NW}/dt$ shown in the inset of Figure 3c) is a direct measure of the photo-induced radial currents inside the NW. The spatial inhomogeneity and the different dynamics of the photo-induced effect result from the local doping contrast along the NW.

The relaxation of the photo-induced effect is governed by the transport properties and the electronic structure of the NW segments. A detailed discussion of the different relaxation processes is beyond the scope of this letter. Here, we limit the discussion to the fast initial dynamics which provide an upper limit for the time resolution of our fsPPM setup. Considering that the built-in radial electric field is on the order of several 10 kV cm$^{-1}$ for heavily doped wires[28], we assume a drift velocity of the photoexcited carriers as high as the saturation velocity in InP, which is $7 \cdot 10^6$ cm s$^{-1}$ [31]. With a wire radius of 15 nm, we expect a drift time of approximately 200 fs, which agrees well with the observed ten-to-ninety rise times of 140 fs and 230 fs of p- and n-segment, respectively. Hence, we interpret the fast initial dynamics as direct measure of radial photocurrent in the nanowire, and conclude that the observed dynamics reflect the carrier dynamics and is not limited by the temporal resolution of our instrument, which according to simulations is expected to be less than 50 fs in the imaging mode[14]. These results demonstrate the feasibility of fsPPM as a novel approach for probing ultrafast currents on the nanoscale with fs temporal resolution.

**Femtosecond LEED**

We further want to discuss the suitability of our setup to study ultrafast structural dynamics in low-dimensional materials by fsLEED. Very recently, Gulde et al. demonstrated the capability of low-energy electrons to study the structural dynamics of a bilayer system on the ps time scale[18]. Here, we introduce an alternative approach for the implementation of time-resolved LEED utilizing the potential of our electron gun design to realize very short propagation distances of the focused beam on μm length scales, therefore minimizing temporal broadening of the electron pulse. The capability of our setup to record high quality LEED patterns of monolayer samples is shown by focusing the electron beam onto single layer graphene suspended over a lacey carbon film[32]. Figure 4a) shows a diffraction pattern recorded in transmission at $d = 500$ μm and 650 eV electron energy exhibiting the six-fold symmetry of the two-dimensional hexagonal lattice of graphene[33]. Noteworthy, even for monolayer samples, diffraction patterns of very high quality can be recorded at very low electron dose rate ($< 1$ e$^-$ Å$^{-2}$ s$^{-1}$) owing to the high scattering cross section of sub-keV electrons[34]. Hence, the implementation of fsLEED for studying structural dynamics in single- and few-layer systems is clearly favorable compared to conventional high energy femtosecond electron diffraction[35].

To study the structural dynamics of such two-dimensional materials after photoexcitation with an ultrashort laser pulse, electron pulses with a length significantly below one picosecond at the sample position are desirable in the diffraction mode. In Figure 4c) the expected full width at half maximum (FWHM) electron pulse duration $\tau_{FWHM}$ is plotted as a function of tip-sample distance for different electron energies, where the focusing condition is adjusted to provide a constant spatial resolution in the diffraction patterns corresponding to a transverse coherence length of ~30 nm (described in more detail in the Supplementary Section III.b). The pulse duration decreases sub-linearly with shorter propagation length, $\tau_{\text{FWHM}} \propto d^{\gamma}$, with $\gamma \approx 0.83$, which can be explained by the distance-dependent reduced inhomogeneity of the acceleration field at the apex in the diffraction mode[14]. So far, the shortest possible distances in the diffraction mode are ~150 μm, restricted by vacuum breakthrough at the electron lens, limiting the electron pulse duration to ~300 fs, see Figure 4c). Future improvements of the lens design should allow distances as close as 20 μm, i.e. distances comparable to the imaging mode, pushing the time resolution of diffraction experiments to the 100 fs range.

We also calculate the electron spot size at the sample and compare it to the experiment. Owing to the absence of space charge and due to the confined emission area, the electron pulses can be focused down to a few μm on the sample, as shown in Figure 4b), where we plot the radially averaged profile revealing a spot size $\rho_{FWHM}$ of 1.4 μm of the focused electron beam. The calculated FWHM spot size, plotted in Figure 4d), linearly decreases with the tip-sample distance down to a few μm, where the slope $\alpha = \Delta\rho_{\text{FWHM}}/\Delta d$, i.e. the beam divergence, depends on the tip voltage according to $\alpha \propto (U_{\text{tip}})^{-1/2}$, reflecting our assumption of constant coherence in the diffraction pattern. Small deviations between simulation and measurement can be due to differences in the probability distributions used for the emission statistics (see Supplementary Section III.b) and due to slightly different focusing conditions. Ultimately, such small electron spot sizes avoid spatial averaging over large domains with multiple crystal orientations, providing an ultrafast structural probe with single-crystal selectivity on μm length scales.

## Discussion

We realized a novel approach for femtosecond point projection microscopy and diffraction using low-energy electron pulses photo-generated from a metal nanotip. We demonstrated the excellent capability of fsPPM for nanoscale imaging of small electric fields around semiconductor nanowires with femtosecond time resolution. In general, fsPPM enables direct spatiotemporal probing of ultrafast processes on nanometer dimensions in the near-surface region of nanostructures, such as ultrafast carrier dynamics and currents, dynamics of interfacial fields as well as ultrafast plasmonics. Ultimately, taking advantage of the high sensitivity of sub-keV femtosecond electron pulses combined with the magnification provided by PPM, our approach potentially allows the investigation of ultrafast phenomena on length scales down to the molecular level[36]. In addition to real space imaging, low-energy electron pulses are ideal probes for studying structural dynamics of 2D crystalline materials on the femtosecond time scale by time-resolved diffraction. Using a nanotip as miniaturized electron gun for fsLEED allows to reduce the electron propagation length to the 100 µm range and to minimize temporal broadening to the 100 fs range. Combining the high surface sensitivity of low-energy electrons with femtosecond time resolution, fsLEED will reveal real-time information on structural dynamics and energy transfer processes in monolayer 2D materials and inorganic[37] as well as organic[38] composite heterostructures thereof.

## Methods

### Setup

The setup is operated by two different laser systems depending on the specific application. For generation of photoelectrons from the tip, a part of the laser output is focused on the tip to a 3-4 µm spot size ($1/e^2$ radius), with the polarization along the tip axis. For time-resolved pump-probe measurements, the second output part is focused onto the sample under an angle of 45°. The arrival time between the electron probe and the optical pump pulse is varied by an optical delay stage integrated in the pump arm (a detailed sketch of the setup is shown in the Supplementary Section I). The interferometric autocorrelation in Figure 1e) was measured at 80 MHz repetition rate with 5 fs pulses and a fluence of 0.14 mJ cm$^{-2}$, with the collimated electron beam at 400 eV electron energy and a copper grid as anode at a distance of ~1 mm. The fsPPM data was measured at 1 MHz repetition rate with 16 fs pulses, with a fluence of 0.7 mJ cm$^{-2}$ focused on the tip and 0.2 mJ cm$^{-2}$ to pump the NWs. An integration time of 2 s was used for each projection image, and the data is averaged over 10 subsequent scans for every delay point. Temporal overlap in Figures 3a)-c) is defined by the empirical multi-exponential fit to the data, see Supplementary Section II.b. For the diffraction data, 5 fs pulses at 80 MHz repetition rate were focused on the tip at a fluence of 0.22 mJ cm$^{-2}$, and diffraction patterns are recorded with an integration time of 0.5 s and averaged over 100 frames. Nanotips with 20-100 nm radii are electrochemically etched from 150 µm polycrystalline tungsten wire. The outer surface of a ceramic tube with an inner (outer) diameter of 200 µm (500 µm) was coated with 100 nm chromium as electron lens. The tip is centered inside the tube and protrudes ~150 µm from the lens. Two additional electrostatic lenses are installed behind the sample to collimate the large diffraction angles obtained in LEED on the plane MCP screen. In the imaging mode these lenses are switched off. A piezo-driven 10-axis positioning system is used for precise alignment of the electron gun and sample inside the laser focuses and relative to each other. All experiments are performed under ultrahigh vacuum conditions ($10^{-10}$ mbar).

### Simulations

The electron pulse duration and spot size at the sample in the fsLEED mode are simulated by classically calculating the single electron trajectories between tip and sample assuming radial symmetry around the tip axis. For the weak field regime in the case of multiphoton photoemission, we can neglect the effect of the optical laser field on the propagation. Gaussian distributions are assumed for the initial electron energy, the emission point along the tip apex as well as the initial electron momentum. More information on the simulations and detailed numbers are given in the Supplementary Section III.a.

**Samples**

InP nanowires with axial p-i-n doping structure are grown as described in reference[24] and mechanically transferred to a gold substrate with a regular pattern of 2 µm holes. Graphene samples are purchased from reference[32] and used without any subsequent treatment.

**Acknowledgements**

We thank M. Borgström and A. Mikkelsen for providing the nanowire samples and helpful discussions. We thank A. Melnikov and A. Alekhin for access to their laser system and S. Kubala for technical assistance. M. M. acknowledges support from the Leibniz Graduate School *Dynamics in new Light* (DinL).

**Author contributions**

R.E. initiated the project; all authors conceived and designed the experiments; M.M. and A.P. constructed the experimental apparatus; M.M. performed the experiments; M.M. and A.P. analyzed the data and performed the numerical simulations; all authors discussed the results and co-wrote the paper.

**Corresponding authors**

Email: m.mueller@fhi-berlin.mpg.de (M.M.) or ernstorfer@fhi-berlin.mpg.de (R.E.).

**Competing financial interests**

The authors declare no competing financial interests.

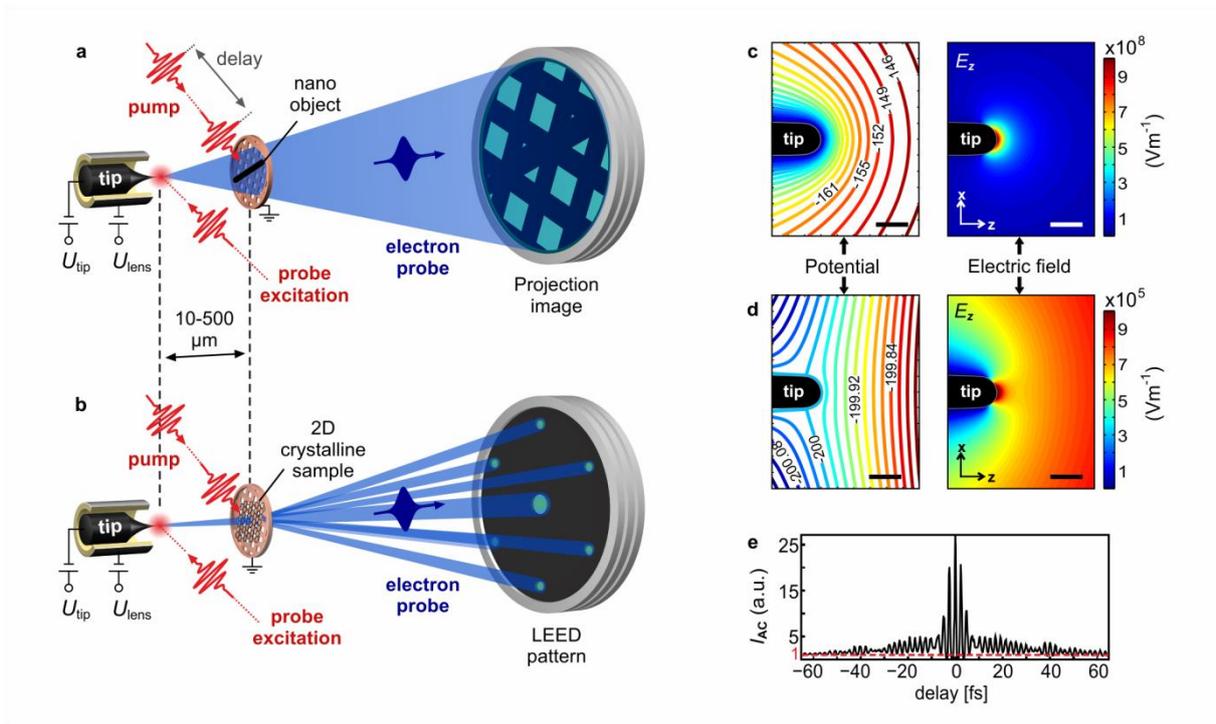

**Figure 1: Setup for time-resolved low-energy electron imaging and diffraction.**
Photoelectrons, generated from a nanotip by an ultrashort laser pulse, are accelerated towards
the sample positioned several μm away from the tip for either (**a**) point projection microscopy
of nanoobjects (divergent electron beam), or (**b**) low-energy electron diffraction of
2-dimensional crystalline samples (collimated beam). A pump laser pulse, variably delayed
from the electron probe, photo-excites the sample for time-resolved experiments. An
electrostatic lens is used to switch from the divergent imaging mode (**c**, curved potential lines
and strong inhomogeneous field $E_z$) to the collimated diffraction mode (**d**, flattened potential
and reduced electric field $E_z$), each at a tip voltage $U_{tip} = -200$ V, but different lens voltages
$U_{lens,im} = -200$ V and $U_{lens,dif} = -730$ V, respectively. Temporally confined electron emission
is verified by measuring the interferometric autocorrelation photocurrent $I_{AC}$ from the tip,
revealing a $3^{rd}$-order emission process (**e**).

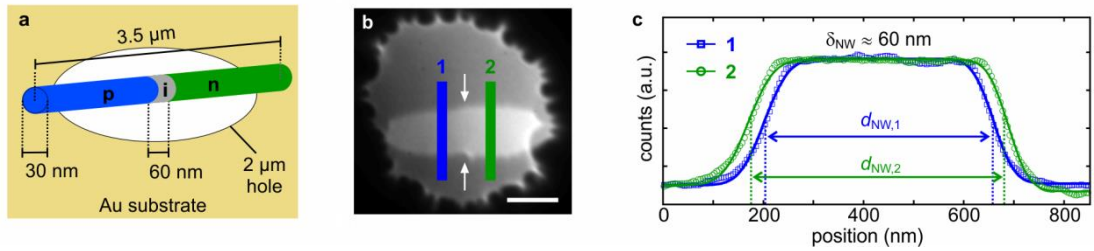

**Figure 2: Point projection microscopy of axially doped nanowires.** InP nanowires (radius 15 nm, length 3.5 µm) with p-i-n axial doping profile and 60 nm i-segment in the center are spanned across 2 µm holes in a gold substrate (**a**). Instead of being a real shadow image of the objects shape, projection images are strongly influenced by local fields surrounding the NW, which becomes apparent by the bright NW projection recorded in constant current (field emission) mode at a tip voltage of -90 V (**b,** scale bar 500 nm). Additionally, a spatial inhomogeneity of the projected diameter along the NW with a step of $\delta_{NW} \approx 60$ nm from the left to the right side of the NW center (marked by the white arrows in (**b**)) is observed (**c**). This corresponds to a potential difference in the 100 meV range and a difference in the radial field around the NW on the order of a few MV m$^{-1}$, as found by simulations (more information on the analysis of the NW diameter is found in the Supplementary Section II and on the simulations in the Supplementary Section III.a).

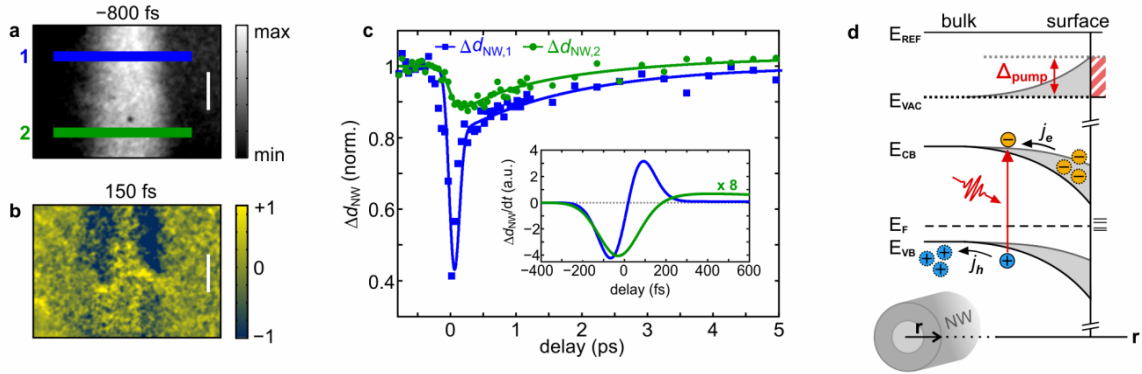

**Figure 3: Femtosecond imaging of ultrafast photocurrents in InP NWs.** (**a**) Projection image of the same NW as in Figure 2b) recorded in pulsed fsPPM mode at negative time delays. Photoecitation by an ultrashort laser pulse leads to a transient, spatially inhomogeneous change of the projected NW diameter (**b**, normalized difference plot). (Data recorded at 70 eV electron energy, scale bars 500 nm). Different dynamical behavior and amplitudes of the transient diameter change $\Delta d_{NW}$ are observed for the two segments along the NW (**c**), where an empirical three-exponential function was fitted to the data. Both segments show a fast initial photo-induced effect with ten-to-ninety rise times in the p- and n-segments of 140 fs and 230 fs, respectively, followed by multi-exponential decay on the fs-to-few ps time scale. As $\Delta d_{NW}$ is directly proportional to the transient electric field change, the derivate $\Delta d_{NW}/dt$ plotted in the inset in (**c**) is a direct measure of the instantaneous photocurrent inside the NW. Surface states cause effective radial doping leading to band bending at the NW surface as sketched in (**d**), where r is the radial coordinate, causing a radial photocurrent of electrons, $j_e$, and holes, $j_h$, after photoexcitation. This leads to a pump-induced transient shift $\Delta_{pump}$ of the conduction band edge $E_{CB}$ and valence band edge $E_{VB}$, and hence a shift of the vacuum level $E_{vac}$ (red shaded area), compared to the reference level $E_{ref}$ (given by the environment), with the magnitude of the shift depending on the specific band bending and doping level.

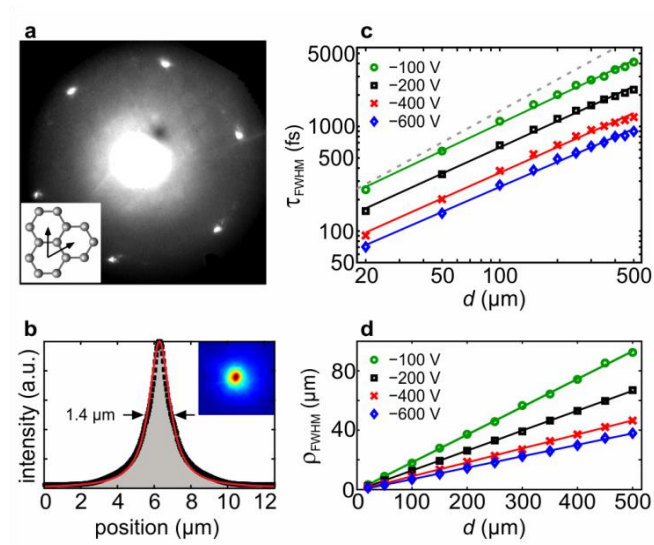

**Figure 4: LEED of free-standing monolayer graphene with fs electron pulses.** LEED pattern of monolayer suspended graphene recorded in transmission at a tip-sample distance of 500 µm and 650 eV electron energy (**a,** inset: hexagonal lattice of graphene). Due to the confined emission area and small propagation distances, the pulsed electron beam can be collimated down to a spot size of 1-2 µm (FWHM) on the sample (**b**), shown here for $d = 200$ µm. The electron pulse duration $\tau_{FWHM}$ in (**c**) is obtained by the FWHM of the arrival time distribution of single electron wave packets for distances $d$ between 20 and 500 µm and electron energies from 100 to 600 eV. A sub-linear dependence $\tau_{FWHM} \propto d^{\gamma}$ with $\gamma \approx 0.83$ is observed ($\gamma = 1$ for the dashed line). Equivalently, the dependence of the electron spot size $\rho_{FWHM}$, defined as the FWHM of the radial position distribution at the sample position, is plotted in (**d**), which is in good agreement with the experimental observations. The dependence on the tip voltage results from the underlying focusing conditions. Further details to the simulations can be found in the Supplementary Section III.b.

# Femtosecond electrons probing currents and atomic structure in nanomaterials

# Supplementary Information

Melanie Müller, Alexander Paarmann, and Ralph Ernstorfer

*Fritz-Haber-Institut der Max-Planck-Gesellschaft, Faradayweg 4-6, D-14195 Berlin, Germany*

## I.   Experimental setup

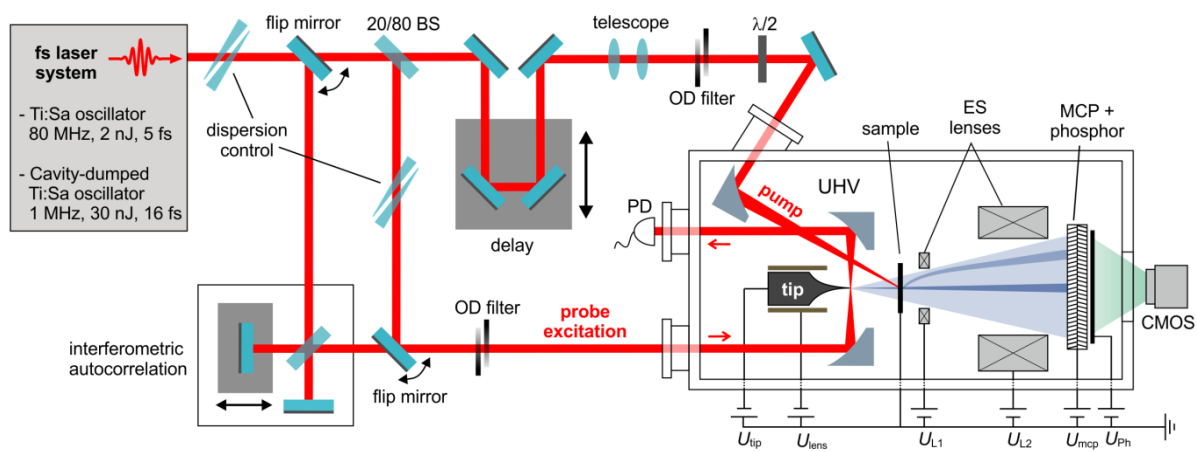

**Figure S1: Experimental setup.** Detailed description is found in the main text.

The setup for femtosecond point projection microscopy (fsPPM) and low-energy electron diffraction (fsLEED) is shown in Figure S1. Two fs laser systems are used: First, an ultra-broadband 800 nm Ti:Sa oscillator running at 80 MHz repetition rate, providing 5 fs pulses with ~2 nJ pulse energy. Second, a cavity-dumped 800 nm Ti:Sa oscillator with variable repetition rate up to 2 MHz delivers 16 fs pulses with 30 nJ pulse energy. Both laser systems can be alternatively incorporated in the same optical setup. A beam stabilization (not shown in the figure) ensures accurate and reproducible alignment of the laser inside the ultrahigh vacuum (UHV) chamber. The laser output is split into two arms for the optical pump and excitation of photoelectrons from the tip as probe, where an optical delay stage is used to vary the delay between pump and probe. An interferometric autocorrelator can be inserted in the probe arm to measure an interferometric autocorrelation of the photocurrent from the tip. Both laser beams are focused by off-axis parabolic mirrors installed inside UHV. A 4-axis positioning stage provides full position alignment of the tip and tilting along the laser beam direction and is used to position the tip (with the electron microlens attached) inside the laser focus. Full position and angle alignment of the sample is achieved by a hexapod-type 6-axis positioning table. The tip can be moved into the pump focus and localized photoemission from the apex is used to precisely align the pump focus position relative to the original tip position.

Bias voltages $U_{tip}$ and $U_{lens}$ up to -2 kV are applied to the tip and the electrostatic microlens depending on the operation mode. Photoelectrons are accelerated towards the grounded sample and amplified by a microchannel plate detector ($U_{mcp}$) combined with a phosphor screen ($U_{Ph}$). A scientific CMOS camera is used to record the images outside UHV. For fsLEED, two electrostatic (ES) lenses at positive bias voltages $U_{L1}$ and $U_{L2}$ are installed behind the sample to reduce the size of the diffraction pattern in order to fit onto the MCP screen.

## II. Data analysis

### a) Projection image analysis

The DC projection image of the p-i-n NW in Figure 2b) is analyzed by taking line profiles at different positions along the NW, see Figure S2.a, and fitting a double error function to the data. The projected NW diameter increases from the substrate contacts at the hole edges towards the NW center (indicated by the white dashed line in Figure S2.a), as plotted in Figures S2.b and S2.c, and saturates close to the center where the i-segment is expected. Noticeably, we observe a constant difference $\delta_{NW}(\Delta x) = d_{NW,1}(\Delta x) - d_{NW,2}(\Delta x)$ at each position $\Delta x$ away from the NW center of $\delta_{\text{NW}}(\Delta x) = 60$ nm, see Figure S2.d. This inhomogeneity clearly indicates different surface fields on both sides of the NW, as expected e.g. for different doping types.

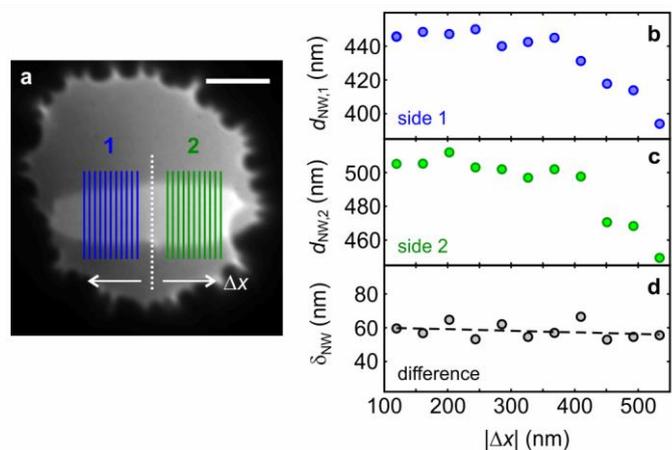

**Figure S2: DC image analysis.** Analysis of the projected diameter along the NW (as indicated by the lines in (**a**)) reveals a constant difference between the left (**b**) and right (**c**) side from the NW center of $\delta_{NW} = 60$ nm (**d**) at all positions $\Delta x$ away from the center towards the hole edges.

**b) Analysis of the time-resolved data**

For each delay frame, the projected width of the nanowire $d_{\text{NW}}$ was fitted with a double error function and averaged over line scans, separately in the blue and green regions indicated in Fig. 3a) of the main text. The dynamics of the extracted values as a function of the delay time $\tau$ plotted in Fig. 3c) of the main text were best fitted empirically with three exponentials

$$\begin{aligned} d_{\text{NW}}(\tau) = A_0 + \Theta(\tau - \tau_0) \\ \times \left( A_1 e^{-\gamma_1(\tau-\tau_0)} + A_2 e^{-\gamma_2(\tau-\tau_0)} + A_3 e^{-\gamma_3(\tau-\tau_0)} + A_\infty \right), \end{aligned} \quad (1)$$

convolved with a Gaussian. Here, $\Theta(t)$ is the Heaviside function, $A_n$ and $\gamma_n$ are the amplitudes and decay rates of the different decay contributions, respectively, and $\tau_0$ is the zero time delay. The constant offsets $A_0$ and $A_\infty$ represent the initial value (before pump) and long-lived contribution to $d_{\text{NW}}(\tau)$, respectively.

## III. Numerical simulations

The numerical simulations were performed with a similar approach as described in reference[1]. A finite element method (FEM) is used to model the electrostatic field between the electron gun and the sample, and in the case of PPM, the detector. Propagation of single electron wave packets inside the electrostatic field is simulated classically using a Runge-Kutta algorithm. The shape of the tip apex is modeled by a half sphere with a 15 nm radius and the shaft has an half opening angle of 13.5°.

**a) Simulation of projection images**

To simulate projection images, we calculate the classical single electron trajectories in three dimensions with cartesian coordinates $x = \{x, y, z\}$, see Figure S3, with the nanowire (NW) spanning across a round hole in x-direction and the tip pointing along the z-direction. Hence, we can choose the x-z-plane as symmetry plane to reduce the computational cost.

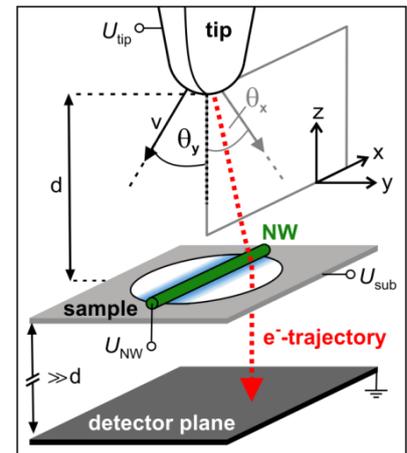

**Figure S3: PPM simulation geometry.** Electrons with initial velocity v and emission angles $\theta_x$ and $\theta_y$ in x- and y-direction, respectively, are accelerated to the sample, and possibly deflected by electric fields in the sample vicinity. Projection images are then evaluated in the distant detector plane.

The sample is modeled by a 200 nm thin metal layer with a 2 µm hole centered on the z-axis.

The NW is formed by a cylinder with radius $r_{NW}$ embedded in the sample. To account for work function variations between the NW and the substrate as well as to the environment (e.g. due to different materials), bias voltages $U_{sub}$ and $U_{NW,0}$ are applied to the sample substrate and the NW, respectively. Additionally, a potential distribution accounting for axial work function variations along the NW, e.g. due to doping effects, can be applied to the NW. To simulate an axial p-i-n doping structure, we model the potential distribution of the NW along the x-direction by

$$U_{NW}(x, x_0, D_x) = U_{NW,0} + U_{NW}^p[1 - \Phi(x, x_0, D_x)] + U_{NW}^n \cdot \Phi(x, x_0, D_x), \qquad (2)$$

with the (cumulative) probability function

$$\Phi(x, x_0, D_x) = \frac{1}{2}\left[1 + \text{erf}\left(\frac{x - x_0}{\sqrt{2}D_x}\right)\right], \qquad (3)$$

the respective potentials $U_{NW}^p$ and $U_{NW}^n$ of the p- and n-doped segments, and with $x_0$ and $D_x$ being the position and width of the i-segment along the x-direction, respectively. In Figures S4.a)-c), examples of the potential $U$ and electric fields $E_x$ and $E_y$ are plotted in the x-y-plane at $z = -20$ μm for a NW with 30 nm radius positioned 20 μm away from the tip, where a potential step of $\Delta U_{pn} = U_{NW}^p - U_{NW}^n = 500$ meV is applied at the NW center with an offset $U_{NW,0} = 1.5$ V. Owing to the nanometer dimensions, electric field strengths of several MV m$^{-1}$ are obtained at such small potentials differences. Even at $U_{NW} = 0$ V, without any potential differences applied to the sample, the electric field strength at the NW surface can reach magnitudes on the MV m$^{-1}$ scale due to the influence of the tip electric field. Ultimately, these fields deflect the electron trajectories close to the NW surface, causing significant lensing effects influencing the projection images.

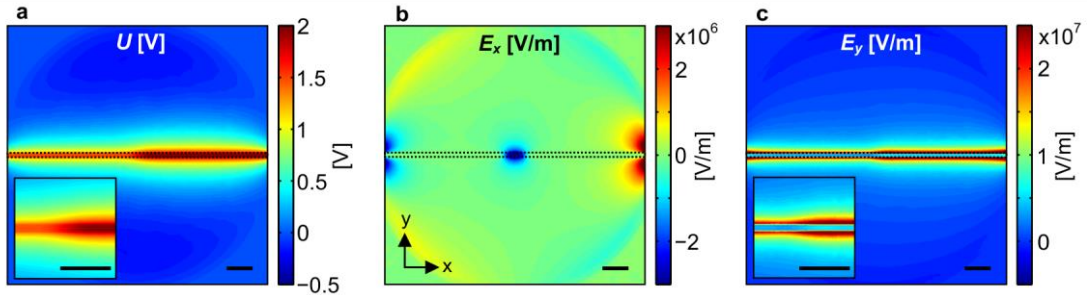

**Figure S4: Potential and electric field of a p-i-n NW.** Potential distribution (**a**) of a p-i-n NW (15 nm radius) with a 500 mV potential step at the NW center and an additional offset of 1.5 V to the substrate . Corresponding electric fields in the x- and y-direction are plotted in (**b**) and (**c**), respectively. All distributions are plotted in the x-y-plane at $z = -20$ μm. All scale bars are 200 nm.

Due to the large computational cost for calculating the projection images, we compute the electron trajectories for a regular grid of emission angles $\theta_x$ and $\theta_y$ in x- and y-direction, respectively, assuming electron emission normal to the tip surface. In addition, a single electron energy is considered since a finite energy distribution has an insignificant effect on the spatial resolution in the projection images compared to other experimental effects like mechanical vibrations and drifts during image acquisition. Projection images are generated by analyzing the arrival positions of all trajectories on the detector plane. Assuming equal emission probability for all trajectories, the image intensity is calculated by phase space mapping between the initial condition and the detector arrival position, integrated over the regular grid of initial conditions.

Figures S5.a) and S5.b) show exemplary projections of a p-i-n NW with constant radius obtained for two different potential distributions, revealing their significance on the projected NW image. The diameter of the projection and its sign, i.e., being a dark or a bright 'shadow', of a certain NW segment depends on its electrostatic potential relative to the substrate, the NW diameter and the tip voltage and distance, respectively. In Figure S5.c) the linear dependence of the projected diameter $d_{NW}$ on the voltage applied to the NW is plotted for various NW radii and two different tip voltages. The threshold voltage $U_{th}$ indicating the transition from dark (positive $d_{NW}$) to bright (negative $d_{NW}$) projections decreases with smaller NW radius and lower tip voltage, respectively, and very thin wires appear bright even a 0 V bias due to the effect of tip electric field.

In conclusion, by numerical simulation of the electron trajectories taking into account all experimental parameters, we can reproduce the recorded projections and relate the observed NW diameters to specific distributions of the potential and electric field at the sample.

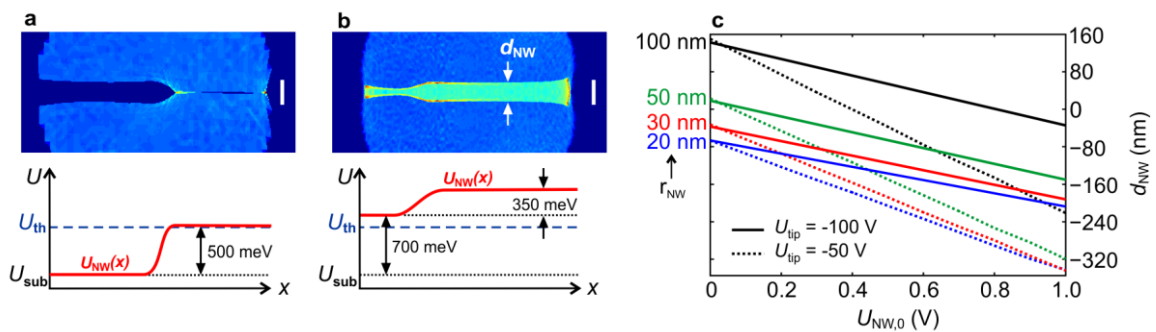

**Figure S5: Projection images and linear field dependence.** Two examples of calculated projection images of a NW with $r_{NW} = 100$ nm at 10 μm distance and -50 V tip voltage are shown for $U_{NW,0} = 0$ V and $\Delta U_{pn} = 0.5$ V (**a**) and for $U_{NW,0} = 0.7$ V and $\Delta U_{pn} = 0.35$ V (**b**), respectively (Scale bars 200 nm). The corresponding potential distributions are sketched below the projection images, with $U_{sub}$ being the potential of the substrate. The transition from dark to bright projections is indicated by the threshold potential $U_{th}$. The dependence of the width and sign of the projected NW diameter $d_{NW}$ on the NW potential is plotted in (c) for NW radii from 20 to 100 nm and two different tip voltages, respectively, revealing a linear dependence on the NW bias.

## Simulation of electron pulse duration and spot size in fsLEED

Assuming cylindrical symmetry, the simulations for the electron pulse duration and spot size in the diffraction mode closely follow the procedure described in reference[1], but additionally including the electron lens. We choose Gaussian distributions for the electron kinetic energy $E$, for the emission angle $\theta_1$ (emission normal to the tip surface), as well as for the momentum distributions at each emission point within and outside the simulation plane, implemented by the angles $\theta_2$ and $\theta_3$, respectively, see Figures S6.a)-c). In particular, the out-of-plane angel $\theta_3$ can be mapped onto the velocity of the electron by $v' = v \cdot \cos(\theta_3)$, effectively reducing the initial electron energy, as the out-of-plane momentum does not affect the arrival time but only induces a precession of the trajectories and their arrival positions around the z-axis (no fields in azimuthal direction due to cylindrical symmetry). Here, the simulations are calculated for a mean energy $E_0 = 0.5$ eV and standard deviations $\sigma_E = 0.25$ eV, $\sigma_{\theta 1} = 10°$, and $\sigma_{\theta 2} = \sigma_{\theta 3} = 30°$ (angles all distributed around zero), adopting the distributions given in reference[2].

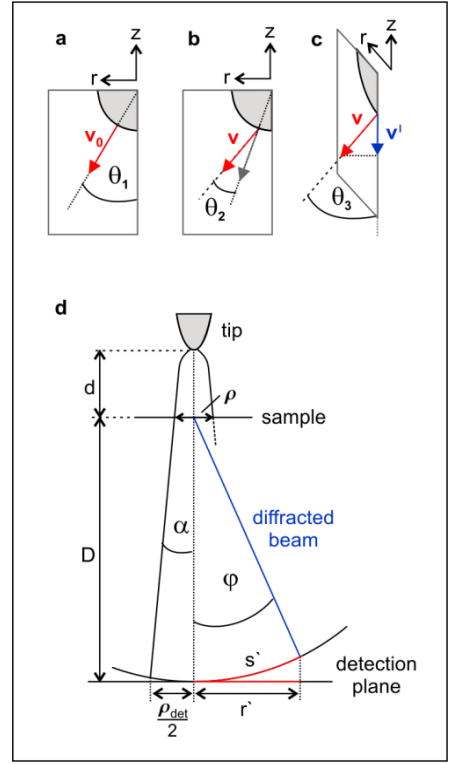

**Figure S6: Simulation geometry for LEED.** Definition of emission angles $\theta_1$ (**a**), normal to the tip surface, and $\theta_2$ (**b**) and $\theta_3$(**c**) accounting for the in- and out-of-plane momentum distributions. (**d**) Sketch of the geometric parameters used to define the focusing condition.

The time-of-flight distribution of the electrons critically depends on the exact field distribution around the tip axis, which changes with tip-sample distance as well as with the tip and lens voltages, respectively. Therefore, we defined an experimentally meaningful focusing condition to compare the results obtained for various distances and electron energies. From the experimental point of view, it is reasonable to assume a constant resolution in the diffraction patterns, i.e., a constant coherence length. In diffraction experiments, the transverse coherence length is usually defined as the ratio between the width of the diffraction spot on the detector, $\rho_{det}$, and its radial position $r'$,[3]

$$l_t = a \frac{r'}{\rho_{det}}, \tag{4}$$

where a is the lattice constant of the investigated sample. For a spherical detector, $r'$ can be defined as the projection of the arc length $s'$ on a planar detection plane, see Figure S6.d), and is proportional to the diffraction angle $\varphi$, i.e., $r' \propto \sin(\varphi)$ in first approximation. According to Bragg's law and the momentum energy relation for non-relativistic electrons with kinetic energy $eU_{\text{tip}}$, we then obtain $r' \propto (U_{\text{tip}})^{-1/2}$. Hence, a constant coherence length for all electron energies requires $\rho_{\text{det}} \propto (U_{\text{tip}})^{-1/2}$ and likewise $\rho \propto (U_{\text{tip}})^{-1/2}$ for the spot size at the sample in the case of field free propagation between sample and detector. In the simulations, this focusing condition is realized by calculating the required electric field strength at the apex which leads to the desired target spot sizes.

The calculations shown in the corresponding letter in Figures 4b) and d) are computed assuming an initial spot size with a standard deviation of $\sigma_\rho = 15$ µm at $U_{\text{tip}} = $ -100 V. The beam divergence $\alpha$ given by the slopes in Figure 4d) show the desired dependence $\alpha \propto (U_{\text{tip}})^{-1/2}$ as plotted in Figure S7. We thus obtain a corresponding spot size on the detector of $\rho_{\text{det}} = 0.37$ mm at a distance of 10 cm. With the Bragg angle of $\varphi = 29.7°$, giving $r' = 0.057$ mm, and using the lattice constant $a = 2.465$ Å of graphene, we obtain a transverse coherence length of $l_{\text{t}} \approx 38$ nm for the above given values. In the same way, we calculate for the coherence length at $U_{\text{tip}} = $ -600 V (with $\sigma_\rho = 6.12$ µm) a value of $l_{\text{t}} \approx 35$ nm, justifying our initial assumptions.

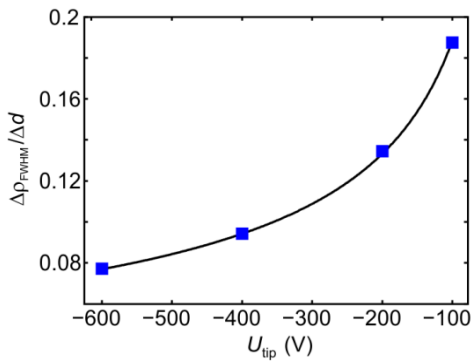

**Figure S7: Voltage dependence of the beam divergence.** Fitted values of the slopes, i.e. the beam divergence, $\Delta\rho_{\text{FWHM}}/\Delta d$ from the data in Figure 4d) in the corresponding letter plotted against the tip voltage.